\documentclass[journal]{IEEEtran}
\usepackage{amsmath,graphicx}
\usepackage{amsfonts}
\newcommand{\eR}{\mathbb{R}} 
\newcommand{\tr}{\operatorname{tr}} 
 
\newcommand{\argmin}{\operatorname{argmin}} 

\usepackage{algorithm}
\usepackage[noend]{algpseudocode}
\usepackage{comment}
\usepackage{url}
\usepackage{comment}
\usepackage{mathtools}
\usepackage{tikz}
\usepackage{float}
\usepackage[justification=centering]{caption}
\usepackage{subcaption}

\usepackage{atbegshi}% http://ctan.org/pkg/atbegshi
\AtBeginDocument{\AtBeginShipoutNext{\AtBeginShipoutDiscard}}
\addtocounter{page}{-1}
\ifCLASSINFOpdf
\else
\fi

\begin{document}
%
% paper title
% Titles are generally capitalized except for words such as a, an, and, as,
% at, but, by, for, in, nor, of, on, or, the, to and up, which are usually
% not capitalized unless they are the first or last word of the title.
% Linebreaks \\ can be used within to get better formatting as desired.
% Do not put math or special symbols in the title.
\title{Graph Regularized Probabilistic Matrix Factorization for Drug-Drug Interactions Prediction}

\author{Stuti Jain,
        Emilie Chouzenoux,~\IEEEmembership{Senior Member,~IEEE,}
        Kriti Kumar,~\IEEEmembership{Member,~IEEE}\\
        and Angshul Majumdar,~\IEEEmembership{Senior Member,~IEEE}}
        % <-this % stops a space
\thanks{S. Jain and E. Chouzenoux are with CVN, Inria Saclay, Univ. Paris Saclay, 91190 Gif-sur-Yvette, France. e-mail: (emilie.chouzenoux@centralesupelec.fr, stuti.jain@inria.fr).}
% <-this % stops a space
\thanks{A. Majumdar and K. Kumar is with Dept. of ECE, IIIT - Delhi, India, 110020. e-mail: (angshul@iiitd.ac.in, kritik@iiitd.ac.in).} 
\thanks{This work received support from the Associate Team COMPASS between Inria and IIIT Delhi. E.C. and S.J. acknowledge support from the European Research Council Starting Grant MAJORIS ERC-2019-STG-850925. }
\maketitle

% As a general rule, do not put math, special symbols or citations
% in the abstract or keywords.
\begin{abstract}
Co-administration of two or more drugs simultaneously can result in adverse drug reactions. Identifying drug-drug interactions (DDIs) is necessary, especially for drug development and for repurposing old drugs. DDI prediction can be viewed as a matrix completion task, for which matrix factorization (MF) appears as a suitable solution. This paper presents a novel Graph Regularized Probabilistic Matrix Factorization (GRPMF) method, which incorporates expert knowledge through a novel graph-based regularization strategy within an MF framework. An efficient and sounded optimization algorithm is proposed to solve the resulting non-convex problem in an alternating fashion. The performance of the proposed method is evaluated through the DrugBank dataset, and comparisons are provided against state-of-the-art techniques. The results demonstrate the superior performance of GRPMF when compared to its counterparts.
\end{abstract}

% Note that keywords are not normally used for peerreview papers.
\begin{IEEEkeywords}
Matrix factorization, Probabilistic matrix factorization, Graph regularization, Drug-drug interaction prediction
\end{IEEEkeywords}

% For peer review papers, you can put extra information on the cover
% page as needed:
% \ifCLASSOPTIONpeerreview
% \begin{center} \bfseries EDICS Category: 3-BBND \end{center}
% \fi
%
% For peerreview papers, this IEEEtran command inserts a page break and
% creates the second title. It will be ignored for other modes.
\IEEEpeerreviewmaketitle

\section{Introduction}
\label{sec:intro}
%Our work can be thought of as a deep extension of probabilistic graph regularized matrix factorization with the same set of antivirals along the rows and columns???
%\\
Drug-Drug Interaction (DDI) refers to the effects of a given drug when taken together with another drug, at the same time. Co-administration of two or more drugs simultaneously can affect the pharmokinetics and/or pharmacodynamics of one or more drugs, which can cause unexpected and even adverse drug reactions \cite{10.1093}. These effects can cause severe injuries to the patients and even be responsible for deaths. Thus, it is necessary to know the DDI for the drugs used in the market, for clinical safety. Knowledge of DDI is also vital for developing new drugs, as well as for repurposing old drugs from clinical and public health perspectives. Pre-clinical identification of DDIs is an ill-posed problem as clinical testing (e.g., in vitro, in vivo, and in populo) is usually conducted on a small group of drugs. Such a process is time and cost intensive. Thus, DDI prediction employing computational approaches that are implementable on a large scale has become a popular research topic in recent years \cite{Roblek}.
%[Roblek, T., Vaupotic, T., Mrhar, A. and Lainscak, M., 2015. Drug-drug interaction software in clinical practice: a systematic review. European journal of clinical pharmacology, 71(2), pp.131-142.].

Computational approaches for DDI prediction can be broadly divided into two categories: (i) Similarity-based approaches, that are based on the similarity of drug information (e.g., chemical structure \cite{ref3}, targets \cite{ref4}, side-effects \cite{ref5}, and (ii) knowledge-based approaches, that employ text mining from scientific literature \cite{ref1}, electronic medical record database \cite{ref2} and the FDA Adverse Event Reporting System \cite{web} to predict DDI. Note that the latter group of approaches do not perform learning per se, as it is mostly a tool to mine clinical findings. Although different machine learning and artificial intelligence models have been used to address the problem \cite{fphar}, DDI prediction remains a challenging problem to address.

In this work, the DDI prediction task is formulated as a matrix completion problem, involving a symmetric matrix with rows/columns corresponding to drugs. We adopt a matrix factorization paradigm, when the sought matrix is defined as the product of two latent factors satisfying some prior knowledge. Our contribution lies in the construction of an original prior well suited to DDI prediction. It incorporates expert knowledge on the DDIs within a graph-based regularization term similar to \cite{DBLP:conf/aaai/StrahlPMK20}, with the aim to favor expected (dis)similarities between drug pairs. This formulation results in the so-called Graph Regularized Probabilistic Matrix Factorization (GRPMF) method, for which we also propose a sounded optimization algorithm relying on modern proximal methods. We evaluate the performance of our method on the DrugBank dataset, and present comparisons against state-of-the-art techniques.  
%The matrix factorization part is responsible for completion; the graph arises from the structural similarity of the drugs. 

The paper is organized as follows. Section II discusses related works on DDI prediction. Section III provides a brief overview of various MF-based methods. Section IV presents our main contribution, that is the proposed GRPMF formulation and the optimization algorithm to resolve it. Section V presents our experimental results and comparisons with the state-of-the-art methods, and finally, Section VI concludes the work.

\section {Related Works}

Clinical trials are time and money consuming. Many DDIs thus remain unknown, mostly because not tested during trials. Machine learning-based methods have been widely investigated to predict the unobserved DDIs. A heterogeneous network-assisted inference (HNAI) method is proposed in \cite{Cheng}, gathering five prediction models (naive Bayes, decision tree, k-nearest neighbor, logistic regression, and support vector machine) to perform DDIs prediction from drug phenotypic, therapeutic, structural, and genomic similarities. The work in \cite{nature1} presents a neural network (NN) based method that proposes a heuristic selection of several drug similarity scores integrated with a nonlinear similarity fusion strategy to obtain abstract features for DDI prediction. Another work \cite{9069437} proposes a semi-supervised learning method for DDI prediction, that uses drug chemical, biological, and phenotype data to calculate the feature similarity of drugs using a regularized least square score minimization. 

Some works address the DDI prediction problem as an edge detection (i.e., link prediction) problem where the edges to infer represent connections between drugs. The work in \cite{journal.pone.0196865} presents both unsupervised and supervised techniques for link prediction using  binary classifiers such as tree, k-nearest neighbors, support vector machine, random forest, and gradient boosting machine based on topological and semantic similarity features to estimate the drug interactions. Another work \cite{journal.pone.0219796} proposes two methods based on NNs and factor propagation over graph nodes, namely, adjacency matrix factorization (AMF) and adjacency matrix factorization with propagation (AMFP) for link prediction for discovering DDIs. 

The superior performance of deep learning (DL) techniques across different domains has triggered the interest in such techniques to estimate drug interactions. The work in \cite{8377981} presents a biomedical resource LSTM (BR-LSTM) that combines biomedical resources with lexical information and entity position information together to extract DDI from the biomedical literature. Note that this model is not DDI prediction per se, but only an automatic tool for mining of information from clinical literature. The work in \cite{ref11} proposes a convolutional mixture density recurrent NN model that integrates convolutional neural networks, recurrent NNs, and mixture density networks for DDI prediction. An autoencoder-based semi-supervised learning algorithm for feature extraction from FDA adverse event reports to identify potential high priority DDIs for medication alerts is presented in \cite{8786248}. Another work \cite{Novel} employs autoencoders and a deep feed-forward network trained on structural similarity profiles (SSP), Gene Ontology (GO) term similarity profiles (GSP), and target gene similarity profiles (TSP) of known drug pairs to predict the effects of DDIs. Due to the black-box nature of the DL models, some work has been done on seeking for explainable DL-based DDI techniques. A comprehensive review of the explainable AI-based techniques to promote the trust of AI models for the critical task of DDI prediction is presented in \cite{VO20222112}.

Recently, some works have utilized Matrix Completion/Factorization (MC/MF) techniques to predict DDIs. Here, given the partially observed DDI matrix, the task is to compute the unobserved interactions between the drugs. Some of the popular generic (not tailored for DDI) MF techniques are (i) singular value decomposition (SVD) \cite{sarwar2000}, (ii) non-negative matrix factorization (NMF) \cite{lee99} and, (iii) probabilistic matrix factorization (PMF) \cite{s2008}. We will present the two later approaches in detail in our next section, as MC/MF constitutes the core of our contribution.%However, these techniques ignore the background information or domain knowledge while making predictions. Hence, for DDI prediction, background information like similarities between drugs based on drug features are incorporated into the matrix factorization frameworks for robust estimation.

In addition to the conventional binary DDI prediction, the work in \cite{NMF} presents an NMF-based approach utilizing drug features for comprehensive DDI prediction. Here, the comprehensive DDI matrix is a signed binary matrix with $+1$ for enhancive drugs, $-1$ for degressive drugs, and $0$ for no drug interactions, respectively, which is rather useful to predict the (positive/negative) behaviors of the interacting drugs. The work in \cite{9310317} presents an attribute supervised learning model probabilistic dependent matrix tri-factorization (PDMTF) approach for adverse DDI prediction. They utilized two drug attributes, molecular structure, side effects, and their correlation to compute the adverse interactions among drugs. The work in \cite{ZHANG201890} introduces a manifold regularized MF (MRMF) technique to predict DDIs using drug similarities based on drug features like substructures, targets, enzymes, transporters, pathways, indications, side effects, and off side effects. 

The publicly available large structured biomedical databases has enabled the use of knowledge graph (KG) based approaches for different applications in the biomedical domain. KGs are used to synthesize large biomedical graphs that map similar drug-related entities in the drug database. The work in \cite{graph-ddi} uses KGs embeddings, namely, RDF2Vec, TransE, TransD, and machine learning algorithms for DDI prediction. A KG NN method (KGNN) that captures the drug and its potential neighborhoods by mining their associated relations in KG for DDI prediction is proposed in \cite{ijcai2020-380}. This method utilizes the drugs’ topological structures in KG for potential DDI prediction. Another work \cite{conv-lstm} utilizes KGs combined with DL techniques for estimating DDIs. This work considers the DDI matrix and KG in the form of learned embeddings (like ComplEx, TransE, RDF2Vec, etc.) as input to the Convolutional Neural Networks (CNN) and Long-Short Term Memory (LSTM) model to predict DDIs. 

In this work, we focus on the MC/MF based framework, as it presents the advantage of being non supervised and highly interpretable. Our contribution is to incorporate expert knowledge within this family of approach, so as to take advantage of the aforementioned progressed in database availability.

\section{Background}
\label{sec:pagestyle}
%In this work, our problem is that of drug-drug interaction. Let $N$ represent the number of drugs in the dataset, the drug-drug interactions is a binary matrix $Y \in \mathbb{R}^{N \times N}$, where $y_{i,j}=1$ if the $i$th drug interacts with the $j$th drug, and $y_{i,j}=0$ if it is unknown if the $i$th and $j$th drug have an interaction or not. Our data has $927$ drugs with $4,09,657$ known interactions. Our aim is to predict the unknown drug interactions.\\

This section presents an overview of MF techniques for MC. We choose here to remain in a generic setting where the matrix to complete is real-valued and rectangular. Note that, for the DDI task, the sought matrix is square symmetric and, in most cases, binary valued, which might lead to simplified formulations. 

%[ ADD A REMARK, SAYING THIS IS A GENERIC REVIEW, BUT IN DDI, Y IS A SQUARE MATRIX.]
% in generic sense.
%[ALSO THIS SECTION MUST DISCUSS THE LIMITATIONS OF THE EXISTING METHODS TO BETTER MOTIVATE THE PROPOSITION]

%[FOR EACH SUBSECTION, YOU NEED TO ADD A REF TO PRIOR WORK]

\subsection{Matrix Completion Problem}

Let us consider the problem of a full matrix $R \in \eR^{N \times M}$ to recover from partially known matrix $Y \in \eR^{N \times M}$.
Let
%Let $X \in \eR^{N \times M}$ be the full matrix to recover. We introduce 
%Let $R \in \eR^{N \times M}$ be the full matrix to recover. We introduce 
\begin{equation}
  \mathcal{D} = \{i \in \{1,\ldots,N\}, j \in \{1,\ldots,M\} \, \text{s.t.} \, (i,j) \, \text{is observed}\} \notag.
\end{equation}
Non observed entries are typically set to zero. The masking of the indexes outside the set $\mathcal{D}$ is modeled through a Hadamard product $\odot$ with a matrix $B \in \{0,1\}^{N \times M}$, such that $B_{ij} = 1$ if $(i,j) \in \mathcal{D}$, and $B_{ij} =0$ otherwise. The partially known matrix $Y$ can be expressed as:
\begin{equation}
    Y = B \odot R.
\end{equation}
The task of matrix completion amounts to recovering the entries of $R$ that do not belong to the set of observed indexes~$\mathcal{D}$.

\subsection{Matrix Factorization (MF)}

MF \cite{KorenBV09} consists of recovering missing entries in matrix $R$ by minimizing the simple least-squares function $\displaystyle{\underset{{R}}{\operatorname{minimize}}}\;   \| Y - B \odot R\|_F^2$ under some specific structural prior constraints on $R$. %:
%The aim of matrix completion is to find $R$ given $Y$.
%We can predict $R$ by minimizing the simple least-squares function:
% \begin{align}
% \displaystyle{\underset{{R}}{\operatorname{minimize}}}\;   \| Y - B \odot R\|_F^2.
% \end{align}
Under MF prior, $R$ is recast as a product of two matrices $U \in \mathbb{R}^{N \times Z}$ and $V \in \mathbb{R}^{Z \times M}$, where $Z \geq 1$ defines a latent space dimension, typically low compared to $(N,M)$. The matrices $U$ and $V$ are inferred by solving:
%For instance, the above mentioned problem can be solved through matrix factorization approach. %One then assumes that $X$ reads as a product of two matrices
%$R$ can be assumed to be a product of two low rank matrices, i.e.,
%$ R = UV $ with $U \in \mathbb{R}^{N \times Z}$ and $V \in %\mathbb{R}^{Z \times M}$, where $Z$ the sought rank for matrix $R$. 
%Hence, we can predict $U$ and $V$ as follows:
\begin{align}
\displaystyle{
\underset{U,V}{\operatorname{minimize}}
}\;   \| Y - B \odot(U V)\|_F^2.
\label{eq:MF}
\end{align}
Subsequently, the complete matrix $R$ is simply recovered by $ R = UV $. Problem \eqref{eq:MF}~is however highly under-determined and extra priors are typically introduced to obtain meaningful solutions. The most widely used being probably the positivity of the entries of the latent factors $(U,V)$, yielding the NMF (nonnegative MF) formulation \cite{NMF_book}. 

Let us discuss related formulations for MC. First, another formulation strategy to impose low rank is to resort to nuclear norm minimization \cite{Recht}. Regularization strategies, based on graph modeling, have been considered in \cite{Gu_SDM} for the MF formulation and in \cite{Mongia_2019} for the nuclear norm formulation. MF models with more than two factors lead to the so-called deep MF approach, investigated for instance in \cite{8683123} in the context of MC. Graph regularized version for deep MC has been proposed in the recent work \cite{Mongia_2022}.

% \textcolor{red}{There are several variants of the basic matrix completion  problem. For example, another approach to address this is via nuclear norm minimization \cite{Recht}. Based on these two approaches, there are several graph regularized versions as well. For example \cite{Gu_SDM} applies it on the factorisation based formulation and \cite{Mongia_2019} employs it on the nuclear norm version. In recent times, matrix completion has been accomplished via deep factorisation as well \cite{8683123}. Graph regularized version for deep matrix factorisation has also been developed in the recent past \cite{Mongia_2022}. All of these studies are only cursorily related to our work and hence, we do not discuss them in detail.} 

%s to the completion/factorization problem. 

\subsection{Probabilistic Matrix Factorization (PMF)}

%The idea of PMF is to solve the completion/factorization problem by setting probabilistic models on $U$ and $V$. 
PMF introduces probabilistic models on the latent factors $U$ and $V$ in the MF formulation \cite{salakhutdinov2008probabilistic}. More precisely:
\begin{itemize}
    \item Each observed entry $(Y_{ij})_{(i,j) \in \mathcal{D}}$ is assumed to follow a Gaussian distribution, with mean $[UV]_{i,j}$ and variance $\sigma^2$ (positive scalar assumed to be known). 
 
    \item Each entry $(U_{iz})_{1 \leq i \leq N, 1 \leq z \leq Z}$ is assumed to follow a Gaussian distribution with zero mean and variance $\sigma_U^2$ (positive scalar assumed to be known).
    
    \item Each entry $(V_{z j})_{1 \leq z \leq Z, 1 \leq j \leq M}$ is assumed to follow a Gaussian distribution with zero mean and variance $\sigma_V^2$ (positive scalar assumed to be known). 
\end{itemize}
%\begin{itemize}
    %\item Each observed entry $(Y_{ij})_{(i,j) \in \mathcal{D}}$ is assumed to follow a Gaussian distribution, with mean equals to the $(i,j)$-th entry of $UV$, and variance equals to $\sigma^2$ (positive scalar assumed to be known). 
    %\item Each entry $(U_{iz})_{1 \leq i \leq N, 1 \leq z \leq Z}$ is assumed to follow a Gaussian distribution with mean equals 0 and variance equals to $\sigma_U^2$ (positive scalar assumed to be known)
    %\item Each entry $(V_{z j})_{1 \leq z \leq Z, 1 \leq j \leq M}$ is assumed to follow a Gaussian distribution with mean equals 0 and variance equals to $\sigma_V^2$ (positive scalar assumed to be known) 
%\end{itemize}
%Searching now for the maximum a posteriori (MAP) estimator of $(U,V)$ given $Y$, associated to the above Gaussian model requires the resolution of:
The maximum a posteriori (MAP) estimator of $(U,V)$ given $Y$, associated with the above model can be obtained by solving:
\begin{multline}
    \underset{U,V}{\text{minimize}} \frac{1}{2 \sigma^2} \| Y - B \odot (UV)\|^2_F
    + \frac{1}{2 \sigma_U^2} \| U \|_F^2 + \frac{1}{2 \sigma_V^2} \| V \|_F^2.
\end{multline}
The minimization with respect to $U$ (resp. $V$) in this formulation amounts to invert a linear system, which can be performed using a conjugate gradient solver. The PMF formulation can be enhanced by incorporating correlated Gaussian distributions, which results in the PMFG formulation described hereafter.
%However, here again, the priors are not sophisticated enough and instability might appear.

\subsection{Probabilistic Matrix Factorization with Graph regularization (PMFG)}

In PMFG, the prior distributions of $U$ and $V$ now include a graph regularization strategy \cite{DBLP:conf/aaai/StrahlPMK20}, which amounts to inferring correlations along the rows (resp. columns) of $U$ (resp. $V$) jointly with the factors $U$ and $V$. These correlations are modeled through two precision matrices, $\Gamma_U \in S_N^{++}$ and $\Gamma_V \in S_M^{++}$, where $S_N^{++} \in \mathbb{R}^{N \times N}$ and $S_M^{++} \in \mathbb{R}^{M \times M}$ denote symmetric positive definite matrices while $S_N$, $S_M$ denote symmetric matrices. The prior is the following:
%Let us denote $S_N$ the space of $N \times N$ symmetric matrices. Moreover, $S_N^+$ (resp. $S_N^{++}$) denotes semi-definite (resp. definite) positive matrices of $S_N$. For PMF with graph regularization, the prior distributions of $U$ and $V$ are more advanced, as we now assume correlations along the lines (resp. columns) of $U$ (resp. $V$). This is modeled by the introduction of two precision matrices, denoted $\Gamma_U \in S_N^{++}$ and $\Gamma_V \in S_M^{++}$.  We consider that :
\begin{itemize}
    \item The columns $u_z \in \eR^N$, $z\in \{1,\ldots,Z\}$, of $U$ are independent realizations of a multivariate Gaussian distribution with zero mean and covariance $C_U = \Gamma_U^{-1}$;
    \item The lines $v_z \in \eR^M$, $z\in \{1,\ldots,Z\}$ are independent realizations of a multivariate Gaussian distribution with zero mean and covariance $C_V = \Gamma_V^{-1}$.
\end{itemize}
The precision matrices $\Gamma_U$ and $\Gamma_V$ are related to Gaussian graphical models associated to the two underlying Gaussian distributions [cite the book Elements-Statistical-Learning-Inference-Prediction], which justifies the name for ``graph'' regularization. Specifically, matrix $\Gamma_U$ (resp. $\Gamma_V$) can be understood as the adjacency matrix of an undirected graph where each edge identifies with two entries of $u_z$ (resp. $v_z$) being correlated, given all the others.

%Searching for the MAP estimator now amounts to solving
The MAP estimate can now be obtained by solving the following:
\begin{multline}
\underset{
\small 
\begin{array}{c}
U,V,\Gamma_U,\Gamma_V
\end{array}
}{\text{minimize}} \frac{1}{2 \sigma^2} \| Y - B \odot (UV)\|^2_F
+ \frac{1}{2} \tr(U^\top \Gamma_U U) \\ + \frac{1}{2}\mathcal{L}( \Gamma_U ) 
    + \frac{1}{2} \tr(V \Gamma_V V^\top) + \frac{1}{2}\mathcal{L}(\Gamma_V ), \label{eq:PMFg1}
\end{multline}
where $\text{tr}(\cdot)$ denotes the trace operation and
\begin{equation}
    (\forall \Gamma_U \in S_N) \quad \mathcal{L}(\Gamma_U) = \begin{cases}
        - \ln \det(\Gamma_U) & \text{if }\Gamma_U \in S_N^{++}\\
        + \infty & \text{otherwise}
    \end{cases}
\end{equation}
with $\text{det}(\cdot)$ the determinant operation.  

PMFG approach provides promising results in \cite{DBLP:conf/aaai/StrahlPMK20}. However, it does not incorporate any physical-oriented knowledge in the sought factors $(U,V)$. Actually, in many applications, such as DDI, expert knowledge is available, that dictates more or less likely correlations among the variables. The aim of this present work is to propose a novel formulation to account for such prior knowledge, within the PMFG paradigm.

% [LIMITATION OF THIS METHOD ?]  
% This method forms the backbone of our proposed GRPMF method for DDI prediction task.

\section{Proposed Graph Regularized Probabilistic Matrix Factorization (GRPMF)}
\label{sec:typestyle}

Let us specify our targeted application. We focus on solving the matrix completion problem arising when predicting interactions between the different drugs, which is the so-called DDI problem. Let $Y \in \mathbb{R}^{N \times N}$ a partially known drug interaction matrix (with unobserved entries set to 0) for $N$ different drugs. The aim is to recover the full drug interaction matrix $R \in \mathbb{R}^{N \times N}$\footnote{In our experimental part, due to specificity of the retained dataset, $R$ is a binary valued matrix. However, our framework holds for any type of real-valued symmetric DDI matrix $R$.}. %be the full interaction matrix to be recovered. 
The proposed method includes expert knowledge within the formulation of PMFG, with the aim of estimating DDIs. Note that unlike general matrix completion problems discussed in previous section, in DDI the matrix to infer is square and symmetric. The sought interaction matrix $R$ can thus be
factored as:
%As our data is drug-drug interaction data which is square and symmetric, there would exist only one factor matrix and it's transpose rather than two separate factor matrices. 
%Mathematically,
\vspace{-5pt}
\begin{equation}
    R = UU^\top.
\end{equation}
Under this new setting, the graphical model is reduced to a single graph with adjacency matrix $\Gamma_U$, and the previously presented PMFG formulation can simply be modified as:
%Hence, (5) can be modified as:
\begin{multline}
\underset{
\small 
\begin{array}{c}
U, \Gamma_U
\end{array}
}{\text{minimize}} \frac{1}{2 \sigma^2} \| Y - B \odot (UU^\top)\|^2_F
+ \frac{1}{2} \tr(U^\top \Gamma_U U)+ \frac{1}{2}\mathcal{L}( \Gamma_U ).  \label{eq:PMFg}
\end{multline}

\subsection{Integrating Expert Knowledge}

As already discussed, in DDI application, one might have prior knowledge about the position of the graph edges (i.e. non zero elements in the precision matrix $\Gamma_U$), thanks to some expert analysis of the database. This prior knowledge is available through an extra symmetric matrix with positive real entries, namely $A_U \in [0,+\infty[^{N \times N}$. For instance, for the DDI prediction task, $A_U$ could result from precomputing the similarity between the $N$ drugs of the dataset in terms of the SIMCOMP (SIMilar COMPound) scores \cite{simcomp}. %[EXPLAIN WHAT WILL BE $A_U$ IN THE CONTEXT OF DDI]
Such matrix can then be used as a structural prior on the sought matrix $\Gamma_U$, so as to remove spurious edges with no physical meaning, and to promote expected ones in the restored graph. Otherwise stated, if for some $(i,j)$ with $i \neq j$, $[A_U]_{ij}$ is large, then $[\Gamma_U]_{ij}=[\Gamma_U]_{ji}$ should be encouraged to be high as well. In contrast, an entry $[A_U]_{ij}$ close or equal to zero should promote the removal of the edge between nodes $i$ and $j$ in the sought graph. 

%Note that we do not need any additional prior on $[\Gamma_U]_{ji}$, as symmetry will be ensured by construction of the optimization method itself.

In order to build a suitable regularization function associated to this new prior, let us introduce the following sets:
\begin{align}
    \mathcal{E}_U & = \{(i,j) \in \{1,\ldots,N\}^2, i \neq j  \, \text{and} \,  [A_U]_{ij} > \tau\} \label{eqE1}\\
\overline{\mathcal{E}}_U & = \{(i,j) \in \{1,\ldots,N\}^2, i \neq j  \, \text{and} \,  [A_U]_{ij} \leq \tau\}  \label{eqE2}
\end{align}
with $\tau\geq0$ a given detection threshold. In order to avoid a tedious non-convex coupling term in the resulting minimization problem, we furthermore introduce the proxy variable $X \in \mathbb{R}^{N \times N}$ that we penalize so as to be close to the sought product $UU^\top$. We thus propose to solve:
\begin{multline}
\underset{\small 
\begin{array}{c}
X,U,\Gamma_U  
\end{array}}{\text{minimize}}
 \frac{1}{2 \sigma^2} \|Y - B \odot X\|^2_F + \frac{1}{2} \tr(U^\top \Gamma_U U) \\  + \lambda_U \sum_{(i,j) \in \overline{\mathcal{E}}_U} | [\Gamma_U]_{ij}| - \lambda_U \sum_{(i,j) \in \mathcal{E}_U} \ln(| [\Gamma_U]_{ij}| + \delta) 
    \\ 
    + \frac{\lambda_R}{2}\|X - UU^\top\|_F^2 + \frac{1}{2}\mathcal{L}( \Gamma_U ) +  \frac{\lambda_U}{2} \| \Gamma_U\|_F^2 .
    \label{eq:PMFgp}
\end{multline}
% \begin{multline}
% \underset{\small 
% \begin{array}{c}
% X \in \mathbb{R}^{N \times M},
% U \in \mathbb{R}^{N \times K},V \in \mathbb{R}^{K \times M} \\
%  \Gamma_U  \in \eR^{N \times N},\Gamma_V \in \eR^{M \times M}
% \end{array}}{\text{minimize}}
%  \frac{1}{2 \sigma^2} \|B \odot (R -  UV)\|^2_F \\ + \frac{1}{2} \tr(U^\top \Gamma_U U)  - \frac{1}{2}\ln \text{det}( \Gamma_U )
%     + \frac{1}{2} \tr(V \Gamma_V V^\top) - \frac{1}{2}\ln \text{det}( \Gamma_V ) \\ + \lambda_U \sum_{i,j \in \overline{\mathcal{E}}_U} | [\Gamma_U]_{ij}| + \lambda_V \sum_{i,j \in \overline{\mathcal{E}}_V} | [\Gamma_V]_{ij}|
%     \\ - \lambda_U \sum_{i,j \in \mathcal{E}_U} \ln(| [\Gamma_U]_{ij}|) - \lambda_V \sum_{i,j \in \mathcal{E}_V} \ln(| [\Gamma_V]_{ij}|)\\
%     + \frac{\lambda_U}{2} \| \Gamma_U\|_F^2 + \frac{\lambda_V}{2} \| \Gamma_V\|_F^2.
%     \label{eq:PMFgp}
% \end{multline}
Hereabove, parameter $\lambda_R>0$ controls the fulfillment of the equality constraint $X = UU^\top$ while parameter $\lambda_U > 0$ controls the regularization imposed on the precision matrix $\Gamma_U$. Our proposed regularization term is made of two parts. We introduce an $\ell_1$ term to promote sparsity on the regions where edges should not appear (i.e. $\overline{\mathcal{E}}_U$), and a log-barrier term, smoothed by $\delta>0$, in regions where edges (i.e., non-zero entries) should be promoted (i.e. $\mathcal{E}_U$). Finally, a quadratic term, that can be viewed as an elastic-net penalty, is added in order to avoid too large values in the entries of the sought covariance matrix.  

\subsection{Optimization Algorithm}
Problem \eqref{eq:PMFgp} is highly non-convex, as it is commonly the case in MF formulations. We propose to use an alternating optimization strategy to solve it. Let $ F(X,U,\Gamma_U)$ denote the loss function presented in \eqref{eq:PMFgp}. Starting from a given initialization $(X^0,U^0,\Gamma_U^0)$, for every iteration $k \in \mathbb{N}$, the parameters of the algorithm are updated as:
\begin{equation}
    \begin{cases}
    X^{k+1} = \underset{X \in \mathbb{R}^{N \times N}}{\argmin}  F(X,U^k,\Gamma_U^k),\\
    U^{k+1} = \underset{U \in \mathbb{R}^{Z \times Z}}{\argmin} F(X^{k+1},U,\Gamma_U^k),\\
    \Gamma_U^{k+1} = \underset{\Gamma_U \in S_N}{\argmin} F(X^{k+1},U^{k+1},\Gamma_U).
    \end{cases}
    \label{algo0}
\end{equation}
Such procedure ensures the monotonical decrease of function $F$. We explicit hereafter each update. Let us remark that, for variables $U$ and $\Gamma_U$, the subproblems remain non-convex. The proposed subroutines only amounts to finding a stationary point for each, which might not be a global minimum. No numerical instabilities were observed in our experiments though. 

\subsubsection{Update of $X$}
The matrix $X$ is updated as:
\begin{multline}
    X^{k+1} = \underset{X \in \mathbb{R}^{N \times N}}{\argmin} \frac{1}{2 \sigma^2} \|Y - B \odot X \|^2_F + \frac{\lambda_R}{2}\|X - U^k (U^k)^\top\|_F^2.
\end{multline}
This is a strictly convex quadratic problem, whose solution satisfies the following optimality condition:
\begin{equation}
    \frac{1}{\sigma^2} Y + \lambda_R U^k (U^k)^\top =   \frac{1}{\sigma^2} B \odot X^{k+1} + \lambda_R X^{k+1}.
    \label{eq:upX}
\end{equation}
The above equation is a linear system that can be solved efficiently by conjugate gradient~\cite{10.5555/865018}.%[ADD A REF].

% \textcolor{red}{Positivity on $X \leadsto$ change solver?}
\subsubsection{Update of $U$}
The matrix $U$ is updated as:
\begin{equation}
   U^{k+1} = \underset{U \in \mathbb{R}^{Z \times Z}}{\argmin} 
   \frac{1}{2} \tr(U^\top \Gamma_U^k U) + \frac{\lambda_R}{2} \|X^{k+1} - U U^\top\|_F^2.
   \label{eq:upU}
\end{equation}
The above optimization problem is non-convex and differentiable. An efficient nonlinear conjugate gradient method is proposed in \cite{DUAN2014236} to address it.
% \begin{equation}
%     (\forall U \in \mathbb{R}^{N \times Z} )\quad     g(U) = \frac{1}{2} \tr(U^\top \Gamma_U^k U) + \frac{\lambda_R}{2} \|X^{k+1} - U U^\top\|_F^2.\\
% \end{equation}
% The gradient of this function reads:
% \begin{equation}
% (\forall U \in \mathbb{R}^{N \times Z}) \quad \nabla{g(U)} = \Gamma_U^k U + \lambda_R(2UU^\top U-(X^\top+X)U).
% \end{equation}

\subsubsection{Update of $\Gamma_U$}

The matrix $\Gamma_U$ is updated using:
\begin{multline}
\Gamma_U^{k+1} =  \underset{\Gamma_U \in S_N}{\argmin}\,
\text{tr}((U^{k+1})^\top \Gamma_U U^{k+1}) +\mathcal{L}(\Gamma_U) + \lambda_U  \| \Gamma_U\|_F^2 \\
+ 2 \lambda_U \sum_{(i,j) \in \overline{\mathcal{E}}_U} | [\Gamma_U]_{ij}|-2 \lambda_U \sum_{(i,j) \in \mathcal{E}_U} \ln(| [\Gamma_U]_{ij}| + \delta)
\end{multline}
or, equivalently,
\begin{equation}
    \Gamma_U^{k+1} = \underset{\Gamma_U \in S_N}{\argmin}\, f(\Gamma_U) + g(\Gamma_U)
\end{equation}
with
\begin{equation}
    (\forall \Gamma_U \in S_N) \quad f(\Gamma_U) = \text{tr}((U^{k+1})^\top \Gamma_U U^{k+1}) +\mathcal{L}(\Gamma_U) 
\end{equation}
and
\begin{multline}
    (\forall \Gamma_U \in S_N) \quad g(\Gamma_U) = 
    2 \lambda_U \sum_{(i,j) \in \overline{\mathcal{E}}_U} | [\Gamma_U]_{ij}|\\
    -2 \lambda_U \sum_{(i,j) \in \mathcal{E}_U} \ln(| [\Gamma_U]_{ij}| + \delta) + \lambda_U \| \Gamma_U\|_F^2.
\end{multline}
The minimization of $f+g$ does not have a close form solution and hereagain, an inner solver is required. Function $f$ is convex, differentiable on its domain $S_N^{++}$, while $g$ is non-convex, non-differentiable. Luckily, the latter is separable over each of the entries of $\Gamma_U$, that is:
\begin{equation}
        (\forall \Gamma_U \in S_N) \quad g(\Gamma_U) = \sum_{1 \leq i,j \leq N} g_{ij}( [\Gamma_U]_{ij})
\end{equation}
with, for every $(i,j) \in \{1,\ldots,N\}^2$,
\begin{multline}
\label{gij}
    (\forall \omega \in \mathbb{R}) \\
    g_{ij}(\omega) = 
    \begin{cases}
    2 \lambda_U |\omega| + \lambda_U \omega^2 & \text{if } (i,j) \in \overline{\mathcal{E}}_U,\\
    - 2 \lambda_U \ln(|\omega| +\delta) + \lambda_U \omega^2 & \text{if } (i,j) \in \mathcal{E}_U.\\
    \end{cases}
\end{multline}

We thus opt for running $L \geq 1$ iterations of a proximal gradient algorithm \cite{attouch:hal-00790042}, %[ADD A REF], 
initialized using the previous value $\Gamma_U^k \in S_N^{++}$ (by construction). This reads as follows:
\begin{equation}
    \begin{array}{l}
    \Gamma_U^{(0)} = \Gamma_U^k\\
    \text{For }\ell=1,2,\ldots,L\\
   \qquad \widetilde{\Gamma}_U^{(\ell)} = \Gamma_U^{(\ell)}
    - \theta^{(\ell)} \nabla f(\Gamma_U^{(\ell)})\\
    \qquad \Gamma_U^{(\ell+1)} = \text{prox}_{\theta^{(\ell)} g} (\widetilde{\Gamma}_U^{(\ell)})\\
     \Gamma_U^{k+1} = \Gamma_U^{(L)}.
    \end{array}
    \label{eq:upGammaU}
\end{equation}
Hereabove, $(\theta^{(\ell)})_{1 \leq \ell \leq L}$ is a sequence of positive stepsizes obtained through a suitable backtracking strategy, so that all iterates remain in the (open) domain of $f$. Moreover, $\text{prox}$ denotes the proximity operator, for which a definition in the non convex case can be found in \cite{attouch:hal-00790042}.

Hereafter, we provide the expression for the gradient of $f$ and the proximity operator of $g$. First, 
\begin{equation}
    (\forall \Gamma_U \in S_N^{++}) \quad 
    \nabla f(\Gamma_U) = U^{k+1} (U^{k+1})^\top - \Gamma_U^{-1},
\end{equation}
while it is not defined for non definite positive matrices of $S_N$. Second, due to the separability of function $g$, we have, for any $\theta>0$ \cite{bauschke:hal-01517477},
\begin{equation}
    (\forall \Gamma_U \in S_N) \quad 
\text{prox}_{\theta g} (\Gamma_U) = 
\left(
\text{prox}_{\theta g_{ij}} ([\Gamma_U]_{ij})
\right)_{1 \leq i,j \leq N}.
\end{equation}
The expression for the proximity operator of each term $g_{ij}$, defined in \eqref{gij}, depends if $(i,j) \in \mathcal{E}_U$ or not. 

For $(i,j)  \in \overline{\mathcal{E}}_U$, $g_{ij}$ is a convex, proper, lower semicontinuous function on $\mathbb{R}$. Its proximity operator is thus uniquely defined, and it reads:

\begin{align}
(\forall \omega \in \mathbb{R}) \;  \text{prox}_{\theta g_{ij}}(\omega) & = \underset{\xi \in \mathbb{R}}{\argmin} \left( \frac{1}{2}(\xi - \omega)^2 \notag \right. \\
& \left.\qquad \qquad \qquad + \theta (2 \lambda_U |\xi|  + \lambda_U \xi^2) \notag \right)\\
        &= \text{prox}_{(1 + 2 \lambda_U \theta)^{-1} |\cdot|}\left( (1 + 2 \lambda_U \theta)^{-1} \omega \right), \notag\\
        &=  \mathcal{S}_{(1 + 2 \lambda_U \theta)^{-1}}\left( (1 + 2 \lambda_U \theta)^{-1} \omega \right),
\end{align}
with $\mathcal{S}_{\tau}$ the soft thresholding operator with parameter $\tau>0$:
\begin{equation}
(\forall u \in \mathbb{R}) \quad
    \mathcal{S}_{\tau}(u) = \text{sign}(u) \max(0,|u|-\tau).
\end{equation}

For $(i,j)  \in \mathcal{E}_U$, $g_{ij}$ is non-convex. Its proximity operator reads:
\begin{align}
    (\forall \omega \in \mathbb{R}) \quad
    \text{prox}_{\theta g_{ij}}(\omega) & = \underset{\xi \in \mathbb{R}}{\argmin} \left( \frac{1}{2}(\xi - \omega)^2 \right. \notag\\
    & \qquad \left.+ \theta (- 2 \lambda_U \ln(|\xi|+\delta) + \lambda_U \xi^2) \right),\\
    & = \text{prox}_{\frac{2\theta\lambda_U}{1 + 2 \theta \lambda_U} \zeta}\left(\frac{1}{1 + 2 \theta \lambda_U} \omega\right),
\end{align}
% & = \text{prox}_{(1 + 2 \lambda_U \theta)^{-1} \zeta}\left( (1 + 2 \lambda_U \theta)^{-1} \omega \right),
with $\zeta: \omega \mapsto - \ln(|\omega|+\delta)$. Function $\zeta$ is non-convex. It is lower semi-continuous, and lower bounded by polynomial $\omega \mapsto - (\omega^2 + \delta$), thus its proximity operator exists though it is not uniquely defined (i.e., it is set-valued). Let $\tau>0$, and $\overline{\omega} \in \mathbb{R}$. Set $\varphi: \omega \mapsto \tau \zeta(\omega) + \frac{1}{2}(\omega - \overline{\omega})^2$. The operator $\operatorname{prox}_{\tau \zeta}$ evaluated at $\overline{\omega}$ is defined as the set of (global) minimizers of $\varphi$. Hereafter, we study $\varphi$, for positive or negative input values, so as to deduce the proximity set of $\tau \zeta$. 

First, for every $\omega > 0$, the first and second order derivatives of $\varphi$ read:
\begin{equation}
    \varphi'(\omega) = - \frac{\tau}{\omega + \delta} + \omega - \overline{\omega}, \label{eq:dphi1}
\end{equation}
and
\begin{equation}
    \varphi''(\omega) = \frac{\tau}{(\omega + \delta)^2} +1. \label{eq:d2phi1}
\end{equation}
Function $\varphi'$ is strictly increasing on $]0,+\infty[$. Moreover, $\lim_{\omega \to 0^+} \varphi'(\omega) = - \frac{\tau}{\delta} - \overline{\omega}$. Thus, $\varphi'$ cancels on $]0,+\infty[$ if $\overline{\omega} > - \frac{\tau}{\delta}$. Canceling \eqref{eq:dphi1} is equivalent to search the roots for the second order polynomial:
\begin{equation}
    - \tau + (\omega - \overline{\omega})(\omega + \delta) 
    =  \omega^2 - (\overline{\omega}- \delta) \omega - \overline{\omega} \delta - \tau.  
\end{equation}
The discriminant of such polynomial is
\begin{align}
    \Delta^+ & = (\overline{\omega}- \delta)^2 + 4( \overline{\omega} \delta + \tau) \\
    & = (\overline{\omega}+ \delta)^2 + 4 \tau, 
\end{align}
which is strictly positive under the condition  $\overline{\omega} > - \frac{\tau}{\delta}$. Two roots exist and read:
\begin{equation}
    \omega_{1,2}^+ = \frac{(\overline{\omega}- \delta) \pm \sqrt{\Delta^+}}{2}.
\end{equation}
There is a sole positive root, equals to:
\begin{equation}
   \omega_{2}^+ = \frac{(\overline{\omega}- \delta) + \sqrt{ (\overline{\omega}+ \delta)^2 + 4 \tau }}{2}.
\end{equation}

Second, when $\omega \in ]-\infty,0[$, we have
\begin{align}
%    (\forall \omega>0) \quad \varphi'(\omega) &= - \frac{\tau}{\omega + \delta} + \omega - \overline{\omega}, \\
   (\forall \omega<0) \quad  \varphi'(\omega) &= \frac{\tau}{- \omega + \delta} + \omega - \overline{\omega} \label{eq:dphi2},%\\
   %\varphi'(0) &= - \frac{\tau}{\delta} - \overline{\omega},
\end{align}
and
\begin{align}
    %(\forall \omega>0) \quad \varphi''(\omega) &=  \frac{\tau}{(\omega + \delta)^2} +1, \\
   (\forall \omega<0) \quad  \varphi''(\omega) &= \frac{\tau}{( \omega - \delta)^2} + 1  .%\\
  % \varphi''(0) &= \frac{\tau}{\delta^2} +1.
\end{align}
Hereagain, $\varphi'$ is strictly increasing. Moreover, $\lim_{\omega \to 0^-} \varphi'(\omega) = \frac{\tau}{\delta} - \overline{\omega}$. Therefore, $\varphi'$ cancels on $]-\infty,0[$ if $\overline{\omega} < \frac{\tau}{\delta}$. 
% Searching for the minimizer(s) of $\varphi$ on $\mathbb{R}$ amounts to:
% \begin{itemize}
%     \item Search for the values $\omega>0$ that cancel \eqref{eq:dphi1};
%       \item Search for the values $\omega<0$ that cancel \eqref{eq:dphi2}; 
%       \item Search, among these values and $0$, the one(s) minimizing $\varphi$.
% \end{itemize}
% First, let us remark that 
Canceling \eqref{eq:dphi2} is equivalent to search the roots for the second order polynomial:
\begin{equation}
     - \tau + (\omega - \overline{\omega})( \omega - \delta) 
    =  \omega^2 - (\overline{\omega}+ \delta) \omega + \overline{\omega} \delta- \tau.
\end{equation}
The discriminant of such polynomial is
\begin{align}
    \Delta^- & = (\overline{\omega}+ \delta)^2 - 4( \overline{\omega} \delta - \tau) =  (\overline{\omega}- \delta)^2 + 4 \tau,
\end{align}
which is strictly positive under the condition $\overline{\omega} < \frac{\tau}{\delta}$. Two roots exist and are given by:
\begin{equation}
    \omega_{1,2}^- = \frac{(\overline{\omega}+ \delta) \pm \sqrt{\Delta^-}}{2}.
\end{equation}
There is a sole negative root, equals to
\begin{equation}
        \omega_{1}^- = \frac{(\overline{\omega}+ \delta) -  \sqrt{(\overline{\omega}+ \delta)^2 - 4( \overline{\omega} \delta - \tau)}}{2}.
\end{equation}

Finally, studying the variations of $\varphi$ on all $\mathbb{R}$ allows us to distinguish three cases:
\begin{itemize}
    \item Case 1: $\overline{\omega} \in ]-\frac{\tau}{\delta},\frac{\tau}{\delta}[$. Then,
    \begin{equation}
        \text{prox}_{\tau \zeta}(\overline{\omega}) = \text{argmin}_{p\in\{ \omega_{1}^-, \omega_{2}^+\}} \varphi(p).
    \end{equation}
    \item Case 2: $\overline{\omega} \leq -\frac{\tau}{\delta}$. Then,
    \begin{equation}
        \text{prox}_{\tau \zeta}(\overline{\omega}) =  \omega_{1}^-.
    \end{equation}
    \item Case 3: $\overline{\omega} \geq \frac{\tau}{\delta}$. Then,
    \begin{equation}
        \text{prox}_{\tau \zeta}(\overline{\omega}) =  \omega_{2}^+.
    \end{equation}
\end{itemize}

\subsubsection{Summarized algorithm}

%[IMPROVE THE TRANSITION !]

%The proposed algorithm~\eqref{algo0} is summarized in Algorithm 1 
%GRPMF [EXPLAIN THE ACRONYM].
%[COMMENT A BIT THE ALGORITHM]

The proposed Graph Regularized Probabilistic Matrix Factorization (GRPMF) method for DDI prediction task is summarized in Algorithm 1. Given the partially observed drug interaction matrix $Y$, we recall that we aim to recover the full drug interaction matrix $R$ by exploiting the drug similarity information captured in $A_U$. Matrices $X$ and $\Gamma_U$ are initialized using identity matrices scaled by a positive value $s^{0}$. The first $Z$ left-singular vectors of $Y$ obtained using singular value decomposition  (SVD) are considered for initialization of $U$. The three unknowns, namely $X$, $U$ and $\Gamma_U$, are updated using the updates defined in the previous subsection, in an iterative manner for $K$ iterations. The final matrix $R$ is recovered using $R =UU^{\top}$.

\begin{algorithm}[H]
	\caption{GRPMF for DDI Prediction}
	\small
	\begin{algorithmic}[1]
% 		\State \textbf{Set}: $L_d,L_t$
        
        \State \textbf{Input:} $Y$, $A_U$
        \State \textbf{Parameters:} $Z, \sigma, s^{0}, \lambda_R, \lambda_U, K, L$.
		\State \textbf{Initialization:} $U^{0}$ (using \texttt{svd}($Y$)), $X^{0} = \Gamma_{U}^{0} = s^{0} \times \mathbb{I}_N$.
		%and
		%\State \textbf{Initialize} 
		%$\mathcal{E}_{U} = \overline{\mathcal{E}_{U}} = 0 \times \mathbb{I}_N$.
		\State Compute $\mathcal{E}_{U}$ and $\overline{\mathcal{E}_{U}}$ using \eqref{eqE1} and \eqref{eqE2}.
		\State \textbf{for} $k=1,2,\ldots,K$ iterations
		%\State
		\State \quad Update $X^{k+1}$ using (\ref{eq:upX});
		\State \quad Update $U^{k+1}$ using (\ref{eq:upU});
		\State \quad Update ${\Gamma_U}^{k+1}$ using (\ref{eq:upGammaU}).
		%\State \quad Update X using the following update (using linear conjugate gradient solver) : %\begin{equation}
        %\frac{1}{\sigma^2} Y + \lambda_R U^k (U^k)^\top =   \frac{1}{\sigma^2} B \odot X^{k+1} + \lambda_R X^{k+1}.
        %\end{equation}
		%\State
		%\State \quad Update  $g(U^k) = \frac{1}{2} \tr((U^k)^\top \Gamma_U^k U^k) + \frac{\lambda_R}{2} \|X^{k+1} - U^k (U^k)^\top\|_F^2$
		%\State
		%\State \quad Update $\nabla{g(U^k)} = \Gamma_U^k U^k + \lambda_R(2U^k(U^k)^\top U^k-(X^\top+X)U^k)$
		%\State 
		%\State \quad \quad \textbf{For} $m=1,2,\ldots,M$ 
		%\State
		%\State  \quad \quad \quad  Update $t^{(m)}$,$U^{(m)}$,$\nabla g(U^{(m)})$, $\beta^{(m)}$, $D^{(m)}$ w.r.t the steps in (16) where $U^{(m+1)} = U^{(m)} + t^{(m)} D^{(m)}$
		%\State 
        %\State \quad \quad \textbf{End For}
        %\State
		%\State \quad \quad \textbf{For} $\ell=1,2,\ldots,L$ 
		%\State \begin{equation}
		%   \widetilde{\Gamma}_U^{(\ell)} = \Gamma_U^{(\ell)}
       % - \theta^{(\ell)} \nabla f(\Gamma_U^{(\ell)})
        %\end{equation}
        %\State
        %\State \begin{equation}
        %\Gamma_U^{(\ell+1)} = \text{prox}_{\theta^{(\ell)} g} (\widetilde{\Gamma}_U^{(\ell)})
        %\end{equation}
        %\State \quad \quad \textbf{End For}
        %\State
		%\State $\Gamma_{U_1}^{k+1} = \Gamma_U^{(L)}$
		%\State
		%\State \quad $R^{k+1} = U^{k+1}{U^{k+1}}^\top$
		%\State
		%\State \quad Set $Diag(R^{k+1}) = 0$
		%\State
		%\State \quad $R^{k+1} = {R^{k+1}}^{+}$
		%\State
		\State \textbf{end}
		\State \textbf{Return}: $R = U^K (U^K)^\top$.% - \text{Diag}(U^K (U^K)^\top)
	\end{algorithmic}
\end{algorithm}

% -------------------------------------------------------------------------

%After updating the parameters using their closed form solutions over few iterations, the final matrix $R$ is recovered.

%[HERE, YOU MUST EXPLAIN BETTER YOUR INITIALIZATION STRATEGY]
%[ALSO, JUSTIFY WHY YOU REMOVE THE DIAGONAL OF R]

\section {Experimental Results}

This section presents our experimental results illustrating the validity of the proposed method. We first introduce the dataset considered for DDI prediction along with the necessary data pre-processing steps carried out. Subsequently, comparison of the proposed GRPMF method against the benchmark algorithms, and an ablation study for GRPMF, are presented.

\subsection{Dataset}
We rely on the DDI data from Stanford University \cite{ref7}  that is approved by the U.S. Food and Drug Administration. It contains $48,514$ interactions from $1,514$ drugs extracted from drug labels and scientific publications \cite{DB, ref6}. We also use information from Our experiments rely on the publicly available drug dataset from the DrugBank dataset gathers $14,315$ drugs along with their KEGG ID or compound ID. In particular, this latter dataset is used to build the expert knowledge for our regularization term. To do so, we rely on the SIMCOMP score that amounts to comparing the chemical structures of the drugs \cite{simcomp}. SIMCOMP computes the similarity of two chemical compounds by counting the number of matched atoms in those atom alignments, using the KEGG ID reference system. We computed the SIMCOMP score using the SIMCOMP search tool with a cutoff of $0.01$ \cite{simcomp_server}. Due to the unavailability of KEGG ID, and thus SIMCOMP scores, for few drugs, and to the non overlap between \cite{ref7} and \cite{DB} drug lists, the final data reduced to $N = 927$ drugs. For this subset of drugs, we made sure that they each have at least $10$ known interactions with other drugs.

%Our experiments rely on the publicly available drug dataset from DrugBank \cite{DB}. 

% Due to the unavailability of KEGG ID for all the drugs, only $3,457$ drugs were retained, for which we computed the SIMCOMP score  using the SIMCOMP search tool with a cutoff of $0.01$ \cite{simcomp_server}. The drug-drug interactions data from Stanford University \cite{ref7} is utilized that is approved by the U.S. Food and Drug Administration. It contains $48,514$ interactions from $1,514$ drugs extracted from drug labels and scientific publications \cite{DB, ref6}. Due to no information on the SIMCOMP score for few drugs, the final data reduced to $927$ drugs %with $4,09,657$ unique interactions 
% that was utilized for the study. It is ensured that this subset of drugs have at least $10$ known interactions with other drugs.

\begin{table*}[!ht]
\centering
\caption{\label{font-table} Quality metrics for all compared methods.}
\begin{tabular}{lllllll}
\hline \textbf{Method} & \textbf{AUPR} & \textbf{AUC} & \textbf{Precision} & \textbf{Recall} & \textbf{F1} & \textbf{Accuracy} \\ \hline
Graph DDI (GNB) & 0.0385  & 0.5018 & 0.9189 & 0.9547 & 0.9359 & 0.9547\\
Graph DDI (LogR) & 0.0385  & 0.5006 & 0.9176 & 0.9579 & 0.9373 & 0.9579\\
Graph DDI (RF) & 0.0384  & 0.5009 & 0.9176 & 0.9579 & 0.9373 & 0.9579\\
\hline
KGNN (Sum)  & 0.1869  & 0.8265 & 0.9466 & 0.8573 & 0.8934 & 0.8573 \\
KGNN (Concat) & 0.1984 & 0.8352 & 0.9466 & 0.8687 & 0.9004 & 0.8687 \\
KGNN (Neighbor) & 0.1076 & 0.7437 & 0.9373 & 0.8167 & 0.8667 & 0.8167 \\
\hline
Conv-LSTM & 0.0381  & 0.4959 & 0.9190 & 0.8621 & 0.8890 & 0.8621\\
%\hline
%MGRNNM &  0.4875 & 0.9170 & 0.9571 & 0.9628 & 0.9507 & 0.9628 \\
%\hline
%MF &  0.2934 & 0.8074 & 0.9491 & 0.9266 & 0.9362 & 0.9266 \\
\hline
GRMF (p = 2, $Z$ = 50, $\lambda_l$ = 0.05, $\lambda_d$ = $\lambda_t$ = 0.3) & 0.4371  &  0.9117 & 0.9620 & 0.9294 &  0.9417 & 0.9294 \\
\hline
%PMFG (p = 2, $Z$ = 20, $\theta$ = 1 , & 37.85  & 85.08 & 91.76 & 95.79 & 93.73 & 95.79\\
% $\sigma$ = 1, $\lambda_U$ = 0, $\lambda_R$ = 0.3, sparam = 0.001)\\
%PMFG (p = 2, $Z$ = 20, $\theta$ = 1 , & 0.0416  & 0.3558 & 0.9176 & 0.9579 & 0.9373 & 0.9579\\
 %$\sigma$ = 1, $\lambda_U$ = 0, $\lambda_R$ = 1, sparam = 0.001)\\
 %PMFG (p = 2, $Z$ = 20, $\theta$ = 1 , & 0.2984  & 0.8340 & 0.9176 & 0.9579 & 0.9373 & 0.9579\\
% $\sigma$ = 1, $\lambda_U$ = 0, $\lambda_R$ = 0.5, sparam = 0.001)\\
PMFG (p = 2, $Z$ = 20, $\theta$ = 0.3, $\lambda_U$ = 0, & 0.3415  & 0.8672 & 0.9495 & 0.9599 & 0.9452 & 0.9599\\
$\lambda_R$ = 1, $\sigma$ = $s^{0}$ = $\delta$ = 0.01)\\
\hline
GRPMF (p = 2, $Z$ = 20, $\theta$ = 0.1, $\lambda_U$ = 0.5,  & \textbf{0.4975}  & \textbf{0.9385} & \textbf{0.9627} & \textbf{0.9617} & \textbf{0.9622} & \textbf{0.9617}\\
$\lambda_R$ = 1, $\sigma$ = $s^{0}$ = $\delta$ = 0.01)\\

%GRPMF (p = 2, $Z$ = 20, $\theta$ = 1 , & \textbf{0.4338}  & \textbf{0.9008} & \textbf{0.9543} & \textbf{0.9620} & \textbf{0.9561} & \textbf{0.9620}\\
%$\sigma$ = 1, $\lambda_U$ = 0.5, $\lambda_R$ = 1, sparam = 0.001)\\

%GRPMF (p = 2, $Z$ = 20, $\theta$ = 1 , & \textbf{51.62}  & \textbf{94.97} & \textbf{96.46} & \textbf{93.51} & \textbf{94.60} & \textbf{93.51} \\
% $\sigma$ = 1, $\lambda_U$ = 0.5, $\lambda_R$ = 1, sparam = 0.001)\\
\hline
\end{tabular}
\label{tab1}
\end{table*}

\begin{figure*}
\centering
\includegraphics[width=14 cm, height=8.5cm]{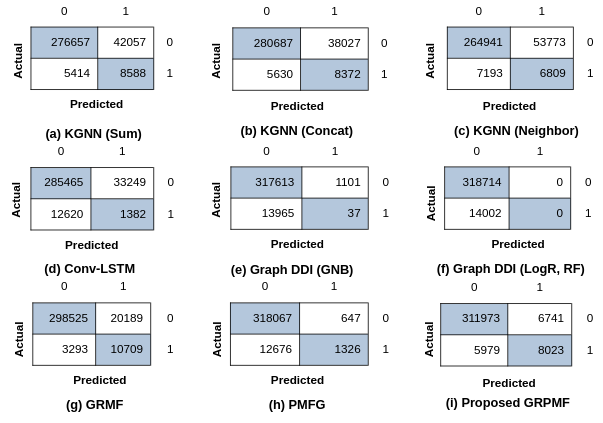}
    %\vspace{-10pt}
\caption{Confusion matrices for all compared methods.}
\label{figure 1}
\end{figure*}

\begin{figure*}
     \centering
     \begin{subfigure}[b]{0.3\textwidth}
         \centering
         \includegraphics[width=\textwidth]{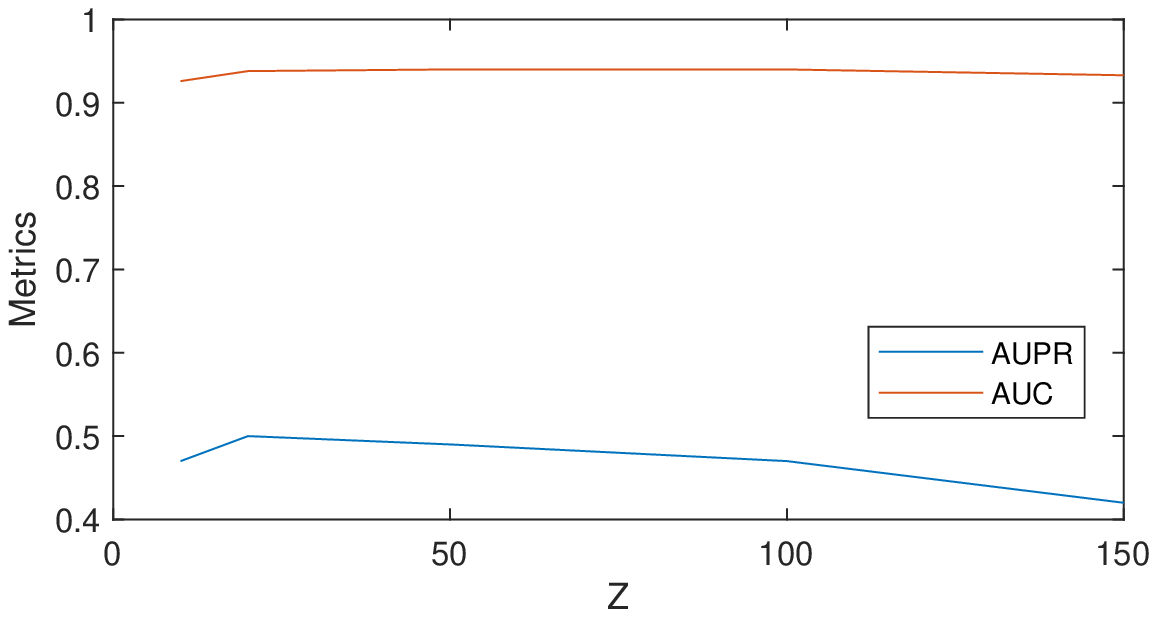}
         \caption{AUPR, AUC vs. $Z$}
     \end{subfigure}
     \hfill
     \begin{subfigure}[b]{0.3\textwidth}
         \centering
         \includegraphics[width=\textwidth]{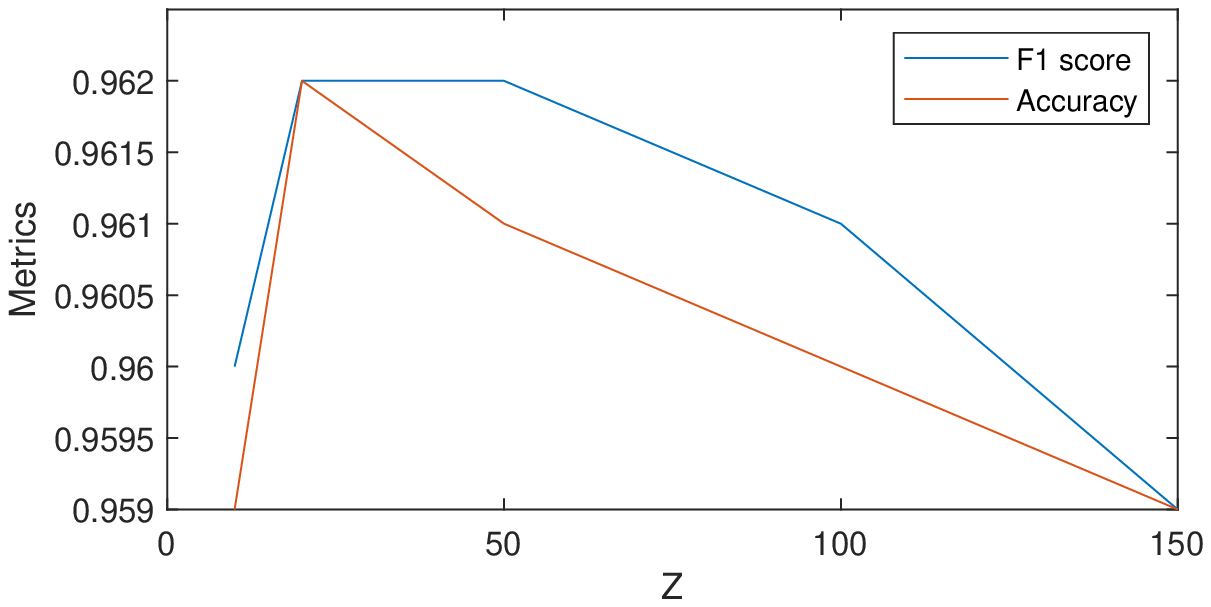}
         \caption{F1 score, Accuracy vs. $Z$}
     \end{subfigure}
     \hfill
     \begin{subfigure}[b]{0.3\textwidth}
         \centering
         \includegraphics[width=\textwidth]{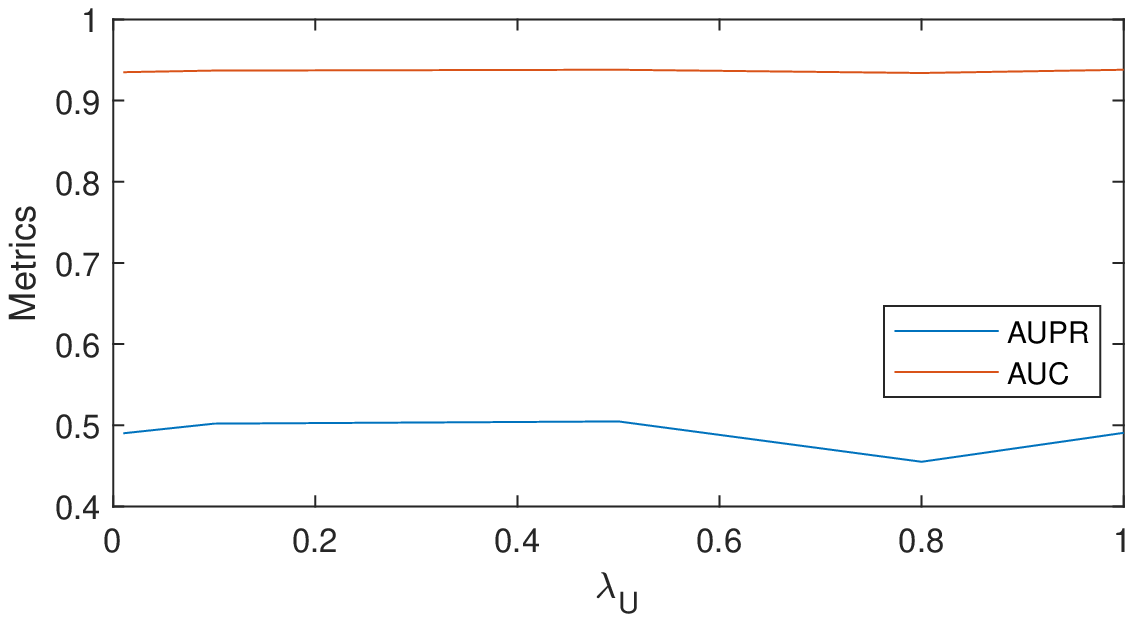}
         \caption{AUPR, AUC vs. $\lambda_U$}
     \end{subfigure}
     \hfill
     \begin{subfigure}[b]{0.3\textwidth}
         \centering
         \includegraphics[width=\textwidth]{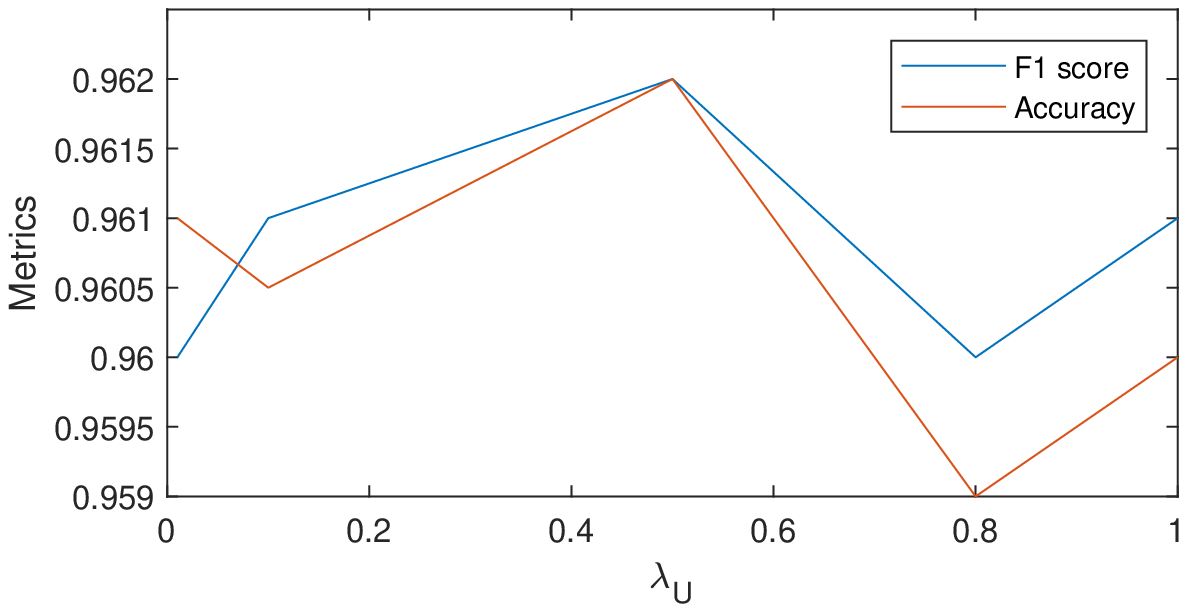}
         \caption{F1 score, Accuracy vs. $\lambda_U$}
     \end{subfigure}
     \hfill
     \begin{subfigure}[b]{0.3\textwidth}
         \centering
         \includegraphics[width=\textwidth]{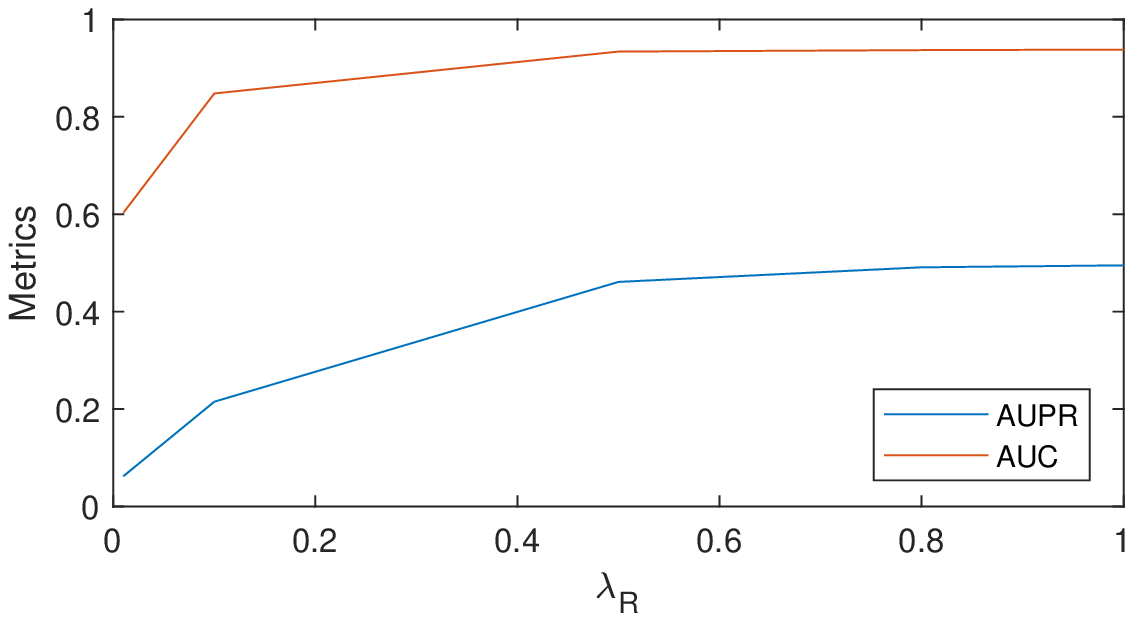}
         \caption{AUPR, AUC vs. $\lambda_R$}
     \end{subfigure}
     \hfill
     \begin{subfigure}[b]{0.3\textwidth}
         \centering
         \includegraphics[width=\textwidth]{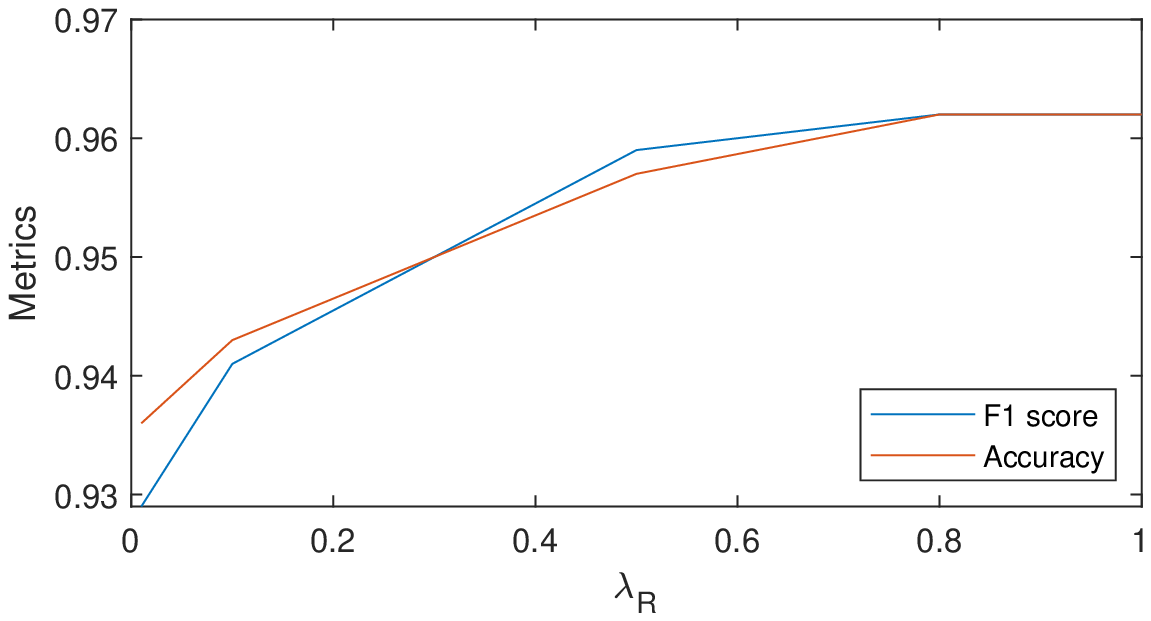}
         \caption{F1 score, Accuracy vs. $\lambda_R$}
     \end{subfigure}
    \caption{Performance of GRPMF with different values of $Z$, $\lambda_U$ and $\lambda_R$}
    \label{fig2}
\end{figure*}

%\subsection{Data Pre-processing}

From our dataset curating, we end up with a {binary matrix} $Y \in \{0,1\}^{N \times N}$, where $y_{i,j}=1$ if drugs $i$ and $j$ are known to interact. An entry $y_{i,j} =0$ {means that no interaction has been reported between drugs $i$ and $j$ so far.}. Our SIMCOMP calculation yields a drug similarity data represented as a symmetric matrix $A_U \in \mathbb{R}^{N \times N}$. Note that the initial SIMCOMP matrix has been sparsified so as to retain only the $p$-nearest neighbours of each drug with the aim to preserve the local geometry of the original data. Such operation promotes that drugs that are close to one another in the original (chemical structure) space should also be close to one another in the learnt (drug interaction) manifold (i.e., local invariance assumption) \cite{ezzat2016drug}. We then compute sets $\mathcal{E}_{U}$, and $\overline{\mathcal{E}_{U}}$ using \eqref{eqE1} and \eqref{eqE2} with $\tau = 0$. 

%The experiments are conducted on the training data that contains all the $927$ drugs with $83$ samples per drug. These $83$ samples from each drug contain $60\%$ samples of known interactions (denoted by 1) and the remaining samples from unknown interactions (denoted by 0). The remaining samples for each drug (belonging to known and unknown interactions) are considered in the test data. The training and test data samples for each drug for both known and unknown interactions are selected randomly. Also, only one pair of interactions are kept that belong from either the upper triangle or the lower triangle of the DDI matrix. This amounts to $76,941$ training samples and $3,32,716$ test samples i.e. $\approx$ 20\% training and $\approx$ 80\% test data.

We then adopt a supervised learning paradigm, as in {\cite{graph-ddi, ijcai2020-380, conv-lstm}.} $20\%$ drug pairs are considered for training (i.e., assumed to be observed), and the remaining $80\%$ are taken for testing (i.e., masked and thus must be estimated). This means that the cardinality of the observed set $\mathcal{D}$ equals $20\% N^2$. Since the number of known interactions (i.e., $1$ entries) are much less than unknown interactions (i.e., $0$ entries) in matrix $Y$, we made sure that the training data contains at most $60\%$ samples of known interactions for each drug. The remaining known interactions for each drug are considered in the test data. The test data comprises of 3,32,716 samples or interactions. Out of this, 3,18,714 samples correspond to class 0 (i.e., no observed drug interaction) and 14,002 correspond to class 1 (i.e., observed drug interaction) that makes the problem very challenging.

% The training and test data for $927$ drugs is curated such that only one pair of interactions are considered that belongs either to the upper triangle or the lower triangle of the DDI matrix. $20\%$ data is considered for training, and the remaining $80\%$ is taken for testing. Since the number of known interactions is very less, it is ensured that the training data contains atmost $60\%$ samples of known interactions for each drug. The remaining known interactions for each drug are considered in the test data. The training and test data samples for each drug for both known and unknown interactions are selected randomly. 

%A simple post-processing is done, consisting in removing the diagonal entries of $R$ that would indicate self interactions (interaction of the drug with itself), and are not of interest in this application.}\textcolor{magenta}{EMILIE SAYS: to say only in the experimental part, in same paragraph where you explain how you threshold negative entries of $R$.}
%\textcolor{magenta}{EMILIE SAYS: this is where you must explain the post-processing, and also that you compute metrics only for true positive.}

\subsection{GRPMF and benchmarks settings}

In all our experiments, GRPMF is applied with the iteration numbers {$K = 10$ and $L = 5$, that appeared enough to reach stability for the inner and outer loops of our algorithm.} As a post-processing, the interaction matrix $R$ recovered using GRPMF is processed by removing the diagonal entries that indicate self interactions (interaction of the drug with itself), and are not of interest in this application. The class labels are assigned using a simple thresholding operation, such that if $|R_{ij}| \leq 0.5$, decision is class $0$, otherwise this is class $1$.

The performance of the proposed method is compared against the following benchmark algorithms %utilizing KGs 
for DDI prediction: 
\begin{itemize}
    \item \textbf{Graph DDI \cite{graph-ddi}}: In this work, the KGs are constructed using different types of embeddings namely, RDF2Vec, TransE and TransD. These embeddings, along with the DDI matrix, are fed one by one to different machine learning techniques like Random Decision Forest (RF), Gaussian Naive Bayes (GNB), and Logistic Regression (LogR) for DDI prediction. Here, only the RDF2Vec embedding vectors with Skip-Gram is retained for comparison as it achieved the best performance in the study \cite{graph-ddi}.
    
    \item \textbf{KGNN \cite{ijcai2020-380}}: Here, the KG results from the drugs' topological structures for DDI prediction. The KG and DDI matrix is fed to Graph Neural Network (GNN). This method focuses on drug neighborhood sampling and aggregates the entities to represent the drugs' potential neighbors in three different ways: (i) Sum, (ii) Concat, and (iii) Neighbor.
   
    \item \textbf{Conv-LSTM \cite{conv-lstm}}: In this method, KG is learned using different embeddings (e.g., ComplEx, TransE, RDF2Vec, etc.) and fed as input to the CNN and LSTM along with the input DDI matrix, to perform DDI prediction. Only ComplEx embedding is considered here for comparison as it gave the best results in \cite{conv-lstm}.  
    
    %\item \textbf{GR1BMC \cite{Mongia_2020}}: This is the graph regularized matrix factorization technique earlier proposed for drug disease association. It has been employed here for DDI prediction where the drug similarity graph is constructed using the SIMCOMP scores for fair comparison with the proposed method. 
    
    %\item \textbf{PMFG}: This is the regular probabilistic matrix factorization with graph regularization method given in (\ref{eq:PMFg}). Results are computed with the same SIMCOMP score for drug similarity using the formulation of the proposed method in (\ref{eq:PMFgp}) with $\lambda_U = 0$.
    
    \item Additionally, comparisons with Graph Regularized Matrix Factorization \textbf{GRMF} \cite {ezzat2016drug} and \textbf{PMFG} (\ref{eq:PMFg}) techniques are presented. The former uses the SIMCOMP drug similarity as an expert knowledge term but, in contrast with our method, it does not infer any graph (i.e., the graph is imposed from the beginning by the user). The method PMFG has been described in the beginning of the paper. Its results are generated using (\ref{eq:PMFgp}) with $\lambda_U = 0$, using the same post-processing procedure as in our method.
\end{itemize}
The hyperparameters for GRMF, PMFG and the proposed GRPMF method are tuned using grid search. The other techniques make use of the hyperparameter values mentioned in their respective works.

Two main performance metrics are used to evaluate the models, namely the Area under the ROC curve (AUC) and the Area under the PR curve (AUPR). In addition, weighted Precision, weighted Recall, weighted F1 score and Accuracy are calculated for the prediction results.

\subsection{Comparison with benchmarks}

Table \ref{tab1} summarizes the values of the performance metrics obtained with the benchmark techniques and the proposed method on the test data. Since we have considered $\approx$ 20\% data for training and the rest for testing, the AUPR of all the models, especially the benchmark models employing deep learning and machine learning models are low. It can be seen from the table that among the learning based models, KGNN (Concat) performs best in terms of AUPR and AUC. The remaining learning based methods have AUC $\approx$ 0.5, indicating no discrimination capability between the classes. They may require more training data for improved performance. The methods based on matrix factorization work much better in this case. Among them, the proposed GRPMF method achieves the best performance across all metrics, with $\approx$ 30\% and $\approx$ 10\% increase in the AUPR and AUC metrics, respectively, over KGNN (Concat) and $\approx$ 6\% increase in the AUPR over GRMF. GRMF reports a better performance over the PMFG method over all the performance metrics. %It even outperforms the PMFG model. 
The high AUPR score of GRPMF indicates that the extra terms in the formulation (\ref{eq:PMFgp}) compared to PMFG (\ref{eq:PMFg}) are able to learn the drug-drug interactions in an effective manner. %The performance of GRMF method on the other hand is comparable to PMFG for most of the metrics except AUC.
%The high score of AUPR compared to PMFG method demonstrates the ability of the GRPMF technique in learning the true drug interactions in a effective manner by the introduction of extra terms in (). %is considerably higher due to............\textcolor{red}{Emilie: Why the proposed method performs better than regular Matrix factorization on graph?}. 

%Out of the total test data of 3,32,716 samples, the support for class 0 (no observed drug interaction) is 3,18,714, and class 1 (observed drug interaction) is 14,002. 
Due to the important class imbalance, the values of accuracy and the weighted metrics, namely, Precision, Recall, and F1 score (given in Table I), are more biased towards the majority class (i.e., class 0 in this case). This is the reason for the difference between the AUPR and AUC metrics compared to all other metrics. Confusion matrices for all the methods are presented in Fig. \ref{figure 1} to provide additional insights on the prediction results. It can be observed that, while KGNN (Figs.  \ref{figure 1} (a)-(c)) performs well in predicting the true interactions, the number of false interactions prediction is also high. The predictions of both Conv-LSTM (Fig.  \ref{figure 1} (d)) and Graph DDI (GNB) (Fig.  \ref{figure 1} (e)) have a low count of true drug interactions. While the other variants, LogR and RF of Graph DDI (Fig.  \ref{figure 1} (f)) are not able to predict the true drug interactions. On the other hand, MF-based techniques appear to have better performance. The number of true positives for the observed drug interactions are more incase of GRMF (Fig.~\ref{figure 1} (h)) compared to PMFG (Fig.~\ref{figure 1} (g)) and hence the former has a higher AUPR and AUC score. While PMFG reports a high number of false negatives, GRMF has more false positives, for known drug interactions. Overall the proposed GRPMF method (Fig.~\ref{figure 1} (i)) reaches the best performance compared to all other methods. The number of true positives are high and false positives are low, both for known and unknown interactions compared to other benchmark methods. 
%Since this method does not use any prior domain knowledge, the performance is low in terms of recall. 
%PMFG method (Fig. 1(h)) has a slightly higher AUPR compared to GRMF but it fails to predict the true drug interactions when the threshold is set to 0.5. 
%In contrast, the proposed GRPMF method (Fig. 1(i)) is able to predict the true drug interactions better with high true positives and low false positives and has the best performance compared to all other methods. 

 \begin{figure}
\centering
\includegraphics[height = 3.5 cm, width = 7.5cm]{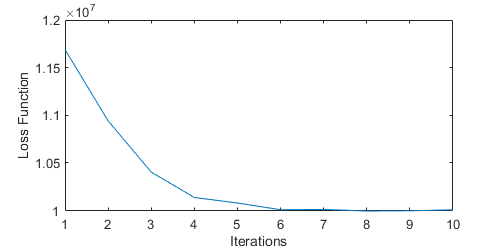}
\caption{Evolution of $F$ along iterations, for GRPMF method.}
\label{fig3}
\end{figure}
%\vspace{-15pt}
 \subsection{Ablation Study}
Three main hyperparameters govern the performance of the proposed GRPMF method: (i) latent dimension $Z$, (ii) expert prior weight $\lambda_U$, and (iii) data fidelity weight $\lambda_R$. We now study the effect of different values of these hyperparameters on the performance of the proposed GRPMF by varying one of the hyperparameters while keeping the others fixed. Figs.~\ref{fig2} (a) and (b) present the plots of all the performance metrics for different values of the $Z$. We observe that AUPR metric is high for $Z = 20$ while the other metrics are comparable for other values of $Z$. Similarly, from Figs.~\ref{fig2} (c) and (d), we observe that $\lambda_U = 0.5$ seems to be an optimal choice across all metrics. Figs.~\ref{fig2} (e) and (f) present the performance variation with the change in $\lambda_R$. Since $\lambda_R$ controls the equality constraint $X = UU^\top$, and as a consequence, the fidelity to the observed data $Y$, the performance is observed to increase with $\lambda_R$, with best value $\lambda_R =1$. The best values for these hyperparameters were considered for generating the results for GRPMF method in Table \ref{tab1}. Finally, Fig.~\ref{fig3} displays the plot of the evolution of the GRPMF loss function $F$ along the iterations for the proposed GRPMF method with the retained settings. It can be seen that $F$ decreases monotonically as expected, and that it reaches stability within a few iterations.

\section{Conclusion}
This paper presents GRPMF, a novel matrix factorization framework for DDI prediction.
Through the introduction of an original regularization strategy, our approach encodes efficiently prior expert knowledge so as to perform jointly DDI prediction and graph similarity inference. Our experimental results obtained with the DrugBank dataset demonstrates the superior performance of our approach compared to deep learning and machine learning-based methods. %\textcolor{red}{Why the proposed method performs better?}
%Experimental results obtained with the DrugBank dataset demonstrate the potential of the proposed technique in estimating complex drug interactions. 
%In our case, the matrix factorization based methods exhibited superior performance compared to the deep learning and machine learning based methods. 
Although the expert knowledge prior in our experiments relied on the SIMCOMP score for drug similarity, other drug similarity features could be considered in future work. Another extension would be to incorporate an explicit constraint favoring binary entries in the sought matrix, which appeared to be beneficial in DDI, and more generally bioinformatics studies \cite{article}. A possible avenue would be to peruse our recent formulation \cite{TALWAR2022103350}, by introducing probabilistic graph modeling terms into it. 

% \textcolor{red}{In DDI and other related bioinformatics problems, the problem is posed as one of binary matrix completion. In that respect, probably using a binary matrix completion framework would be more beneficial \cite{article}. In the recent past, a work proposed graph regularised binary matrix completion as well \cite{TALWAR2022103350}; this work was based on a deterministic framework. In the future, we would like to develop a probabilistic formulation for the same.}

\bibliographystyle{IEEEtran}
\bibliography{references}
\end{document}